\documentclass[reprint,aps,prl,twocolumn,superscriptaddress,showpacs,preprintnumbers,amsmath,amssymb,floatfix,merge]{revtex4-1}
\usepackage{graphicx}  % needed for figures
\usepackage[mathlines]{lineno}% Enable numbering of text and display math
\usepackage{dcolumn}   % needed for some tables
\usepackage{bm}        % for math
\usepackage{amssymb}   % for math
\usepackage{xcolor}
\usepackage{verbatim}
\usepackage{tabularx} 
\usepackage[colorlinks,linkcolor=blue,citecolor=blue,urlcolor=blue]{hyper ref}
\usepackage{array}
\def\gev{GeV/\textit{c}$^2$}

\begin{document}

% The following information is for internal review, please remove them for submission
%\widetext
%\linenumbers
%\leftline{To be submitted to PRL}
%\centerline{\em \small{CDMS INTERNAL DOCUMENT -- NOT FOR PUBLIC DISTRIBUTION}.}
%\centerline{\em \small{WE ARE BREAKING UP INTO SECTIONS FOR NOW. PRL HAS NO SECTIONS.}}

% the following line is for submission, including submission to the arXiv!!
%\hspace{5.2in} \mbox{Fermilab-Pub-04/xxx-E}

\title{Search for Low-Mass WIMPs with SuperCDMS}
%Put affiliations first to force them into alphabetical order
\affiliation{Division of Physics, Mathematics, \& Astronomy, California Institute of Technology, Pasadena, CA 91125, USA} 
\affiliation{Fermi National Accelerator Laboratory, Batavia, IL 60510, USA}
\affiliation{Lawrence Berkeley National Laboratory, Berkeley, CA 94720, USA}
\affiliation{Department of Physics, Massachusetts Institute of Technology, Cambridge, MA 02139, USA}
\affiliation{Pacific Northwest National Laboratory, Richland, WA 99352, USA}
\affiliation{Department of Physics, Queen's University, Kingston, ON K7L 3N6, Canada}
\affiliation{Department of Physics, Santa Clara University, Santa Clara, CA 95053, USA}
\affiliation{SLAC National Accelerator Laboratory/Kavli Institute for Particle Astrophysics and Cosmology, 2575 Sand Hill Road, Menlo Park 94025, CA}
\affiliation{Department of Physics, Southern Methodist University, Dallas, TX 75275, USA}
\affiliation{Department of Physics, Stanford University, Stanford, CA 94305, USA}
\affiliation{Department of Physics, Syracuse University, Syracuse, NY 13244, USA}
\affiliation{Department of Physics, Texas A\&M University, College Station, TX 77843, USA}
\affiliation{Departamento de F\'{\i}sica Te\'orica and Instituto de F\'{\i}sica Te\'orica UAM/CSIC, Universidad Aut\'onoma de Madrid, 28049 Madrid, Spain}
\affiliation{Department of Physics, University of California, Berkeley, CA 94720, USA}
\affiliation{Department of Physics, University of California, Santa Barbara, CA 93106, USA}
\affiliation{Department of Physics, University of Colorado Denver, Denver, CO 80217, USA}
\affiliation{Department of Physics, University of Evansville, Evansville, IN 47722, USA}
\affiliation{Department of Physics, University of Florida, Gainesville, FL 32611, USA}
\affiliation{Department of Physics, University of Illinois at Urbana-Champaign, Urbana, IL 61801, USA}
\affiliation{School of Physics \& Astronomy, University of Minnesota, Minneapolis, MN 55455, USA}
%\affiliation{Physics Institute, University of Z\"{u}rich, Winterthurerstr. 190, CH-8057, Switzerland}

% Now the authors from Tarek's file in CVS
\author{R.~Agnese} \affiliation{Department of Physics, University of Florida, Gainesville, FL 32611, USA}
\author{A.J.~Anderson} \email{Corresponding author: adama@mit.edu} \affiliation{Department of Physics, Massachusetts Institute of Technology, Cambridge, MA 02139, USA}
\author{M.~Asai} \affiliation{SLAC National Accelerator Laboratory/Kavli Institute for Particle Astrophysics and Cosmology, 2575 Sand Hill Road, Menlo Park 94025, CA}
\author{D.~Balakishiyeva} \affiliation{Department of Physics, University of Florida, Gainesville, FL 32611, USA}
\author{R.~Basu~Thakur~} \affiliation{Fermi National Accelerator Laboratory, Batavia, IL 60510, USA}\affiliation{Department of Physics, University of Illinois at Urbana-Champaign, Urbana, IL 61801, USA}
\author{D.A.~Bauer} \affiliation{Fermi National Accelerator Laboratory, Batavia, IL 60510, USA}
\author{J.~Beaty} \affiliation{School of Physics \& Astronomy, University of Minnesota, Minneapolis, MN 55455, USA}
\author{J.~Billard} \affiliation{Department of Physics, Massachusetts Institute of Technology, Cambridge, MA 02139, USA}
\author{A.~Borgland} \affiliation{SLAC National Accelerator Laboratory/Kavli Institute for Particle Astrophysics and Cosmology, 2575 Sand Hill Road, Menlo Park 94025, CA}
\author{M.A.~Bowles} \affiliation{Department of Physics, Syracuse University, Syracuse, NY 13244, USA}
\author{D.~Brandt} \affiliation{SLAC National Accelerator Laboratory/Kavli Institute for Particle Astrophysics and Cosmology, 2575 Sand Hill Road, Menlo Park 94025, CA}
\author{P.L.~Brink} \affiliation{SLAC National Accelerator Laboratory/Kavli Institute for Particle Astrophysics and Cosmology, 2575 Sand Hill Road, Menlo Park 94025, CA}
\author{R.~Bunker} \affiliation{Department of Physics, Syracuse University, Syracuse, NY 13244, USA}
\author{B.~Cabrera} \affiliation{Department of Physics, Stanford University, Stanford, CA 94305, USA}
\author{D.O.~Caldwell} \affiliation{Department of Physics, University of California, Santa Barbara, CA 93106, USA}
\author{D.G.~Cerdeno} \affiliation{Departamento de F\'{\i}sica Te\'orica and Instituto de F\'{\i}sica Te\'orica UAM/CSIC, Universidad Aut\'onoma de Madrid, 28049 Madrid, Spain}
\author{H.~Chagani} \affiliation{School of Physics \& Astronomy, University of Minnesota, Minneapolis, MN 55455, USA}
\author{Y.~Chen} \affiliation{Department of Physics, Syracuse University, Syracuse, NY 13244, USA}
\author{M.~Cherry} \affiliation{SLAC National Accelerator Laboratory/Kavli Institute for Particle Astrophysics and Cosmology, 2575 Sand Hill Road, Menlo Park 94025, CA}
\author{J.~Cooley} \affiliation{Department of Physics, Southern Methodist University, Dallas, TX 75275, USA}
\author{B.~Cornell} \affiliation{Division of Physics, Mathematics, \& Astronomy, California Institute of Technology, Pasadena, CA 91125, USA}
\author{C.H.~Crewdson} \affiliation{Department of Physics, Queen's University, Kingston ON, Canada K7L 3N6}
\author{P.~Cushman} \affiliation{School of Physics \& Astronomy, University of Minnesota, Minneapolis, MN 55455, USA}
\author{M.~Daal} \affiliation{Department of Physics, University of California, Berkeley, CA 94720, USA}
\author{D.~DeVaney} \affiliation{School of Physics \& Astronomy, University of Minnesota, Minneapolis, MN 55455, USA}
\author{P.C.F.~Di~Stefano} \affiliation{Department of Physics, Queen's University, Kingston ON, Canada K7L 3N6}
\author{E.~Do~Couto~E~Silva} \affiliation{SLAC National Accelerator Laboratory/Kavli Institute for Particle Astrophysics and Cosmology, 2575 Sand Hill Road, Menlo Park 94025, CA}
\author{T.~Doughty} \affiliation{Department of Physics, University of California, Berkeley, CA 94720, USA}
\author{L.~Esteban} \affiliation{Departamento de F\'{\i}sica Te\'orica and Instituto de F\'{\i}sica Te\'orica UAM/CSIC, Universidad Aut\'onoma de Madrid, 28049 Madrid, Spain}
\author{S.~Fallows} \affiliation{School of Physics \& Astronomy, University of Minnesota, Minneapolis, MN 55455, USA}
\author{E.~Figueroa-Feliciano} \affiliation{Department of Physics, Massachusetts Institute of Technology, Cambridge, MA 02139, USA}
\author{G.L.~Godfrey} \affiliation{SLAC National Accelerator Laboratory/Kavli Institute for Particle Astrophysics and Cosmology, 2575 Sand Hill Road, Menlo Park 94025, CA}
\author{S.R.~Golwala} \affiliation{Division of Physics, Mathematics, \& Astronomy, California Institute of Technology, Pasadena, CA 91125, USA}
\author{J.~Hall} \affiliation{Pacific Northwest National Laboratory, Richland, WA 99352, USA}
\author{S.~Hansen} \affiliation{Fermi National Accelerator Laboratory, Batavia, IL 60510, USA}
\author{H.R.~Harris} \affiliation{Department of Physics, Texas A\&M University, College Station, TX 77843, USA}
\author{S.A.~Hertel} \affiliation{Department of Physics, Massachusetts Institute of Technology, Cambridge, MA 02139, USA}
\author{B.A.~Hines} \affiliation{Department of Physics, University of Colorado Denver, Denver, CO 80217, USA}
\author{T.~Hofer} \affiliation{School of Physics \& Astronomy, University of Minnesota, Minneapolis, MN 55455, USA}
\author{D.~Holmgren} \affiliation{Fermi National Accelerator Laboratory, Batavia, IL 60510, USA}
\author{L.~Hsu} \affiliation{Fermi National Accelerator Laboratory, Batavia, IL 60510, USA}
\author{M.E.~Huber} \affiliation{Department of Physics, University of Colorado Denver, Denver, CO 80217, USA}
\author{A.~Jastram} \affiliation{Department of Physics, Texas A\&M University, College Station, TX 77843, USA}
\author{O.~Kamaev} \affiliation{Department of Physics, Queen's University, Kingston ON, Canada K7L 3N6}
\author{B.~Kara} \affiliation{Department of Physics, Southern Methodist University, Dallas, TX 75275, USA}
\author{M.H.~Kelsey} \affiliation{SLAC National Accelerator Laboratory/Kavli Institute for Particle Astrophysics and Cosmology, 2575 Sand Hill Road, Menlo Park 94025, CA}
\author{S.~Kenany} \affiliation{Department of Physics, University of California, Berkeley, CA 94720, USA}
\author{A.~Kennedy} \affiliation{School of Physics \& Astronomy, University of Minnesota, Minneapolis, MN 55455, USA}
\author{M.~Kiveni} \affiliation{Department of Physics, Syracuse University, Syracuse, NY 13244, USA}
\author{K.~Koch} \affiliation{School of Physics \& Astronomy, University of Minnesota, Minneapolis, MN 55455, USA}
\author{A.~Leder} \affiliation{Department of Physics, Massachusetts Institute of Technology, Cambridge, MA 02139, USA}
\author{B.~Loer} \affiliation{Fermi National Accelerator Laboratory, Batavia, IL 60510, USA}
\author{E.~Lopez~Asamar} \affiliation{Departamento de F\'{\i}sica Te\'orica and Instituto de F\'{\i}sica Te\'orica UAM/CSIC, Universidad Aut\'onoma de Madrid, 28049 Madrid, Spain}
\author{R.~Mahapatra} \affiliation{Department of Physics, Texas A\&M University, College Station, TX 77843, USA}
\author{V.~Mandic} \affiliation{School of Physics \& Astronomy, University of Minnesota, Minneapolis, MN 55455, USA}
\author{C.~Martinez} \affiliation{Department of Physics, Queen's University, Kingston ON, Canada K7L 3N6}
\author{K.A.~McCarthy} \affiliation{Department of Physics, Massachusetts Institute of Technology, Cambridge, MA 02139, USA}
\author{N.~Mirabolfathi} \affiliation{Department of Physics, University of California, Berkeley, CA 94720, USA}
\author{R.A.~Moffatt} \affiliation{Department of Physics, Stanford University, Stanford, CA 94305, USA}
\author{R.H.~Nelson} \affiliation{Division of Physics, Mathematics, \& Astronomy, California Institute of Technology, Pasadena, CA 91125, USA}
\author{L.~Novak} \affiliation{Department of Physics, Stanford University, Stanford, CA 94305, USA}
\author{K.~Page} \affiliation{Department of Physics, Queen's University, Kingston ON, Canada K7L 3N6}
\author{R.~Partridge} \affiliation{SLAC National Accelerator Laboratory/Kavli Institute for Particle Astrophysics and Cosmology, 2575 Sand Hill Road, Menlo Park 94025, CA}
\author{M.~Pepin} \affiliation{School of Physics \& Astronomy, University of Minnesota, Minneapolis, MN 55455, USA}
\author{A.~Phipps} \affiliation{Department of Physics, University of California, Berkeley, CA 94720, USA}
\author{M.~Platt} \affiliation{Department of Physics, Texas A\&M University, College Station, TX 77843, USA}
\author{K.~Prasad} \affiliation{Department of Physics, Texas A\&M University, College Station, TX 77843, USA}
\author{M.~Pyle} \affiliation{Department of Physics, University of California, Berkeley, CA 94720, USA}
\author{H.~Qiu} \affiliation{Department of Physics, Southern Methodist University, Dallas, TX 75275, USA}
\author{W.~Rau} \affiliation{Department of Physics, Queen's University, Kingston ON, Canada K7L 3N6}
\author{P.~Redl} \affiliation{Department of Physics, Stanford University, Stanford, CA 94305, USA}
\author{A.~Reisetter} \affiliation{Department of Physics, University of Evansville, Evansville, IN 47722, USA}
\author{R.W.~Resch} \affiliation{SLAC National Accelerator Laboratory/Kavli Institute for Particle Astrophysics and Cosmology, 2575 Sand Hill Road, Menlo Park 94025, CA}
\author{Y.~Ricci} \affiliation{Department of Physics, Queen's University, Kingston ON, Canada K7L 3N6}
\author{M.~Ruschman} \affiliation{Fermi National Accelerator Laboratory, Batavia, IL 60510, USA}
\author{T.~Saab} \affiliation{Department of Physics, University of Florida, Gainesville, FL 32611, USA}
\author{B.~Sadoulet} \affiliation{Department of Physics, University of California, Berkeley, CA 94720, USA}\affiliation{Lawrence Berkeley National Laboratory, Berkeley, CA 94720, USA}
\author{J.~Sander} \affiliation{Department of Physics, University of South Dakota, Vermillion, SD 57069, USA}
\author{R.L.~Schmitt} \affiliation{Fermi National Accelerator Laboratory, Batavia, IL 60510, USA}
\author{K.~Schneck} \affiliation{SLAC National Accelerator Laboratory/Kavli Institute for Particle Astrophysics and Cosmology, 2575 Sand Hill Road, Menlo Park 94025, CA}
\author{R.W.~Schnee} \affiliation{Department of Physics, Syracuse University, Syracuse, NY 13244, USA}
\author{S.~Scorza} \affiliation{Department of Physics, Southern Methodist University, Dallas, TX 75275, USA}
\author{D.N.~Seitz} \affiliation{Department of Physics, University of California, Berkeley, CA 94720, USA}
\author{B.~Serfass} \affiliation{Department of Physics, University of California, Berkeley, CA 94720, USA}
\author{B.~Shank} \affiliation{Department of Physics, Stanford University, Stanford, CA 94305, USA}
\author{D.~Speller} \affiliation{Department of Physics, University of California, Berkeley, CA 94720, USA}
\author{A.~Tomada} \affiliation{SLAC National Accelerator Laboratory/Kavli Institute for Particle Astrophysics and Cosmology, 2575 Sand Hill Road, Menlo Park 94025, CA}
\author{S.~Upadhyayula} \affiliation{Department of Physics, Texas A\&M University, College Station, TX 77843, USA}
\author{A.N.~Villano} \affiliation{School of Physics \& Astronomy, University of Minnesota, Minneapolis, MN 55455, USA}
\author{B.~Welliver} \affiliation{Department of Physics, University of Florida, Gainesville, FL 32611, USA}
\author{D.H.~Wright} \affiliation{SLAC National Accelerator Laboratory/Kavli Institute for Particle Astrophysics and Cosmology, 2575 Sand Hill Road, Menlo Park 94025, CA}
\author{S.~Yellin} \affiliation{Department of Physics, Stanford University, Stanford, CA 94305, USA}
\author{J.J.~Yen} \affiliation{Department of Physics, Stanford University, Stanford, CA 94305, USA}
\author{B.A.~Young} \affiliation{Department of Physics, Santa Clara University, Santa Clara, CA 95053, USA}
\author{J.~Zhang} \affiliation{School of Physics \& Astronomy, University of Minnesota, Minneapolis, MN 55455, USA}

\collaboration{The SuperCDMS Collaboration} 
\noaffiliation

\begin{abstract}

We report a first search for weakly interacting massive particles (WIMPs) using the background rejection capabilities of SuperCDMS. An exposure of 577~kg-days was analyzed for WIMPs with mass $<$ 30~\gev, with the signal region blinded. Eleven events were observed after unblinding. We set an upper limit on the spin-independent WIMP-nucleon cross section of $1.2\times10^{-42}$cm$^{2}$ at 8~\gev. This result is in tension with WIMP interpretations of recent experiments and probes new parameter space for WIMP-nucleon scattering for WIMP masses $<$ 6~\gev.

\end{abstract}

\pacs{}
\maketitle

%--- Physics intro ----

Evidence on galactic and cosmological scales strongly indicates that $\sim 80$\% of the matter density of the Universe consists of non-luminous, non-baryonic dark matter, whose particle nature remains unknown \cite{doi:10.1146/annurev-astro-082708-101659}. Weakly interacting massive particles (WIMPs) are one class of theoretically well-motivated candidates for dark matter and may be detectable by searching for keV-scale nuclear recoils in terrestrial detectors \cite{PhysRevD.31.3059}.  Recent excesses of events reported by CDMS~II (Si) \cite{Agnese:2013rvf}, CoGeNT \cite{PhysRevD.88.012002}, CRESST-II \cite{Angloher:2012fk}, DAMA \cite{Bernabei:2010uq}, and possible indirect evidence from gamma rays from the galactic center \cite{PhysRevD.84.123005}, may have been caused by a light WIMP with mass in the 6--30~\gev~range. A variety of theoretical models also favor light WIMPs in this mass range \cite{PhysRevLett.68.741, PhysRevD.79.115016, Falkowski:2011fk, doi:10.1142/S0217751X13300287, Zurek:2013wia, Essig:2010ye, PhysRevD.80.035008, PhysRevLett.110.041302}.

Since light WIMPs produce only low-energy nuclear recoils, experiments optimized for masses $\gtrsim$~30~\gev\ have searched for light WIMPs by lowering their analysis energy thresholds \cite{R21LowT, PhysRevLett.106.131302, PhysRevD.86.051701, PhysRevLett.107.051301}. This additional sensitivity comes with higher background rates because resolution effects degrade particle discrimination at low energies. Following this approach, we analyzed low-energy recoils in the range 1.6--10 keV$_{\rm nr}$ (nuclear-recoil equivalent energy) from the SuperCDMS experiment at the Soudan Underground Laboratory (SUL)~\cite{cdmslitePRL,2012_Rau_TAUP}. Although background discrimination gradually degrades with decreasing event energy, some discrimination can still be achieved using the relative signals measured by the different readout channels on each detector.

%--- Experimental setup and dataset ----
SuperCDMS at Soudan is an upgrade to the Cryogenic Dark Matter Search (CDMS II)~\cite{CDMSScience:2010} with new detector hardware, and is operating in the same location with the same low-radioactivity setup \cite{soudan_setup}. The target consists of fifteen 0.6-kg cylindrical germanium crystals stacked in groups of three to form five towers.  These detectors, known as iZIPs, are instrumented with interleaved ionization and phonon sensors on their flat faces.  From the measured ionization and phonon energy, we derive the recoil energy and the ``ionization yield," the ratio between ionization and recoil energy.  Nuclear recoils, expected from WIMPs, exhibit a reduced ionization yield compared to electron recoils, which are expected from most backgrounds.  The iZIP sensor layout improves the ability to define a fiducial volume in the bulk (fiducialization) compared to the CDMS~II design~\cite{2013iZIPdiscrimination}. The fraction of the total phonon or ionization energy measured by the guard sensors provides radial fiducialization through the ``radial partition" parameter, and the fraction measured by the sensors on each face provides \textit{z} fiducialization through the ``\textit{z} partition" parameter. Such fiducialization rejects events in the peripheral regions of the detectors.  These ``surface events'' often suffer from reduced ionization signal, thus polluting the WIMP signal region.

The SuperCDMS payload has been operating in SUL since March 2012.  The data presented here, recorded between October 2012 and June 2013, are a subset of the ongoing exposure. The seven detectors with the lowest trigger thresholds are used for this search.   The remaining detectors are used to reject events with energy deposition in more than one detector. Consistency tests are used to remove periods of abnormal detector behavior and elevated noise.   After accounting for these losses, the exposure is 577~kg-days. To prevent bias when defining the event-selection criteria, all single-detector hits with recoil energies between a time-dependent noise threshold and 10~keV$_{\rm nr}$, and with ionization energy consistent with nuclear recoils, were removed from study, i.e.~blinded. An exception was made for periods following $^{252}$Cf calibrations, when background rates were higher because of neutron activation of the detectors and their copper housings. This ``open'' dataset constitutes 97~kg-days of exposure that is distinct from the 577 kg-days of data analyzed for WIMPs, and was not used in the final limit calculation or to optimize selection criteria.

%--- calibration and energyscale ----
For the detectors analyzed, the standard deviation of the baseline noise is $\lesssim260$\,eV for summed phonon channels and $\lesssim460$\,eV$_{\rm ee}$ (electron-recoil equivalent) for individual ionization channels. The electron- and nuclear-recoil energy scales are calibrated in a fashion similar to the CDMS II light-WIMP search~\cite{PhysRevLett.106.131302} using $^{133}$Ba and $^{252}$Cf sources respectively. A small ($\lesssim10$\%) variation of the phonon signal gain with the cryostat base temperature, which varied over the range 54--62~mK, is taken into account by the phonon calibration. In each detector, the mean ionization energy of nuclear recoils as a function of total phonon energy, as determined from $^{252}$Cf calibration data, is consistent with, or slightly below, the prediction of Lindhard~\cite{Lindhard, Lewin199687}. A nuclear-recoil band was constructed by accepting events within 3$\sigma$ of the mean ionization energy. Nuclear-recoil equivalent energies are reconstructed from the total phonon energy by subtracting the contribution of Luke-Neganov phonons \cite{neganov, LukeAPhys} corresponding to the mean nuclear-recoil ionization response for the respective total phonon energy.

%--- quality and preselection cuts ----

\begin{figure}[t!!]
\begin{center}
\includegraphics[width=250 pt]{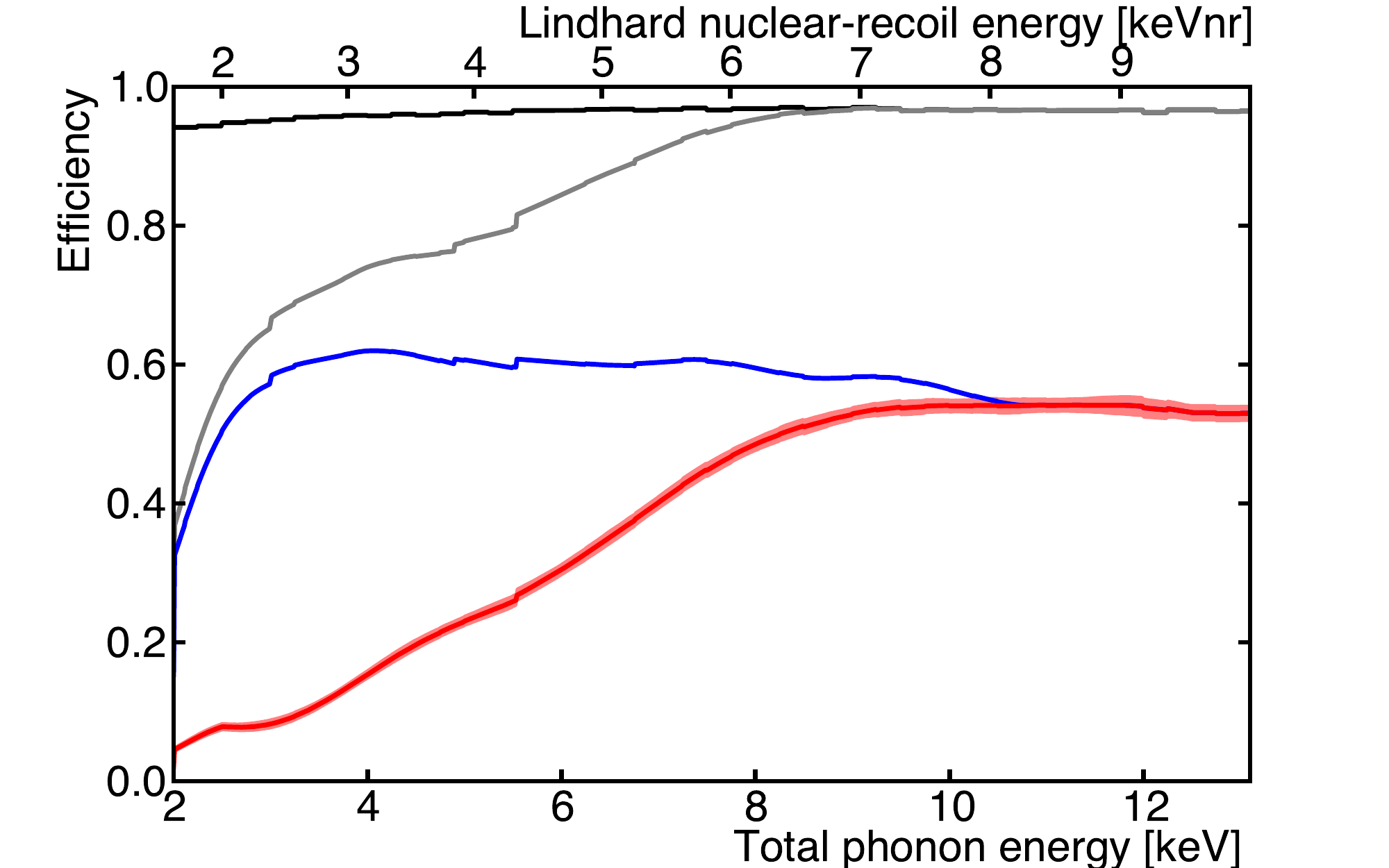}
\caption{Cumulative efficiencies after sequential application of each stage of event selection. From top to bottom, these are data-quality criteria, trigger and analysis thresholds, preselection criteria, and BDT discrimination with 68\% C.L. (stat. + syst.) uncertainty band. The preselection and BDT selection efficiencies are interpolated from measurements in 1~keV bins. Steps are due to time-dependent analysis thresholds for individual detectors. For illustrative purposes, an approximate nuclear-recoil energy scale is provided.}
\label{efficiency}
\end{center}
\end{figure}

Hardware trigger thresholds for each detector were adjusted several times during the WIMP search. For each period of constant trigger threshold, the trigger efficiencies as functions of total phonon energy were measured using $^{133}$Ba calibration data.   The fit results were found to be consistent with, and more precise than, ones obtained using $^{252}$Cf and multiple-hit WIMP-search data.  Analysis thresholds are set to be $1\sigma$ below the energy at which the detector trigger efficiency is 50\% in periods of time for which this quantity is above the noise threshold used in the data blinding, and equal to such threshold otherwise.  The combined efficiency is an exposure-weighted sum of the measured efficiency for each detector and period, shown in Fig.~\ref{efficiency}.

To be selected as WIMP candidates, triggered events had to pass three levels of data-selection criteria: data quality, preselection, and event discrimination. Figure~\ref{efficiency} shows the cumulative efficiency after applying each level of selection criteria and the analysis thresholds.  The first level of criteria (data quality) rejects poorly reconstructed and noise-induced events.  Periods of abnormal noise are removed by requiring that the pre-trigger baseline noise of each event be consistent with normal periods.  Spurious triggers caused by electronic glitches and low-frequency noise in the phonon channels, which populate the low-energy region, were rejected using a pulse-shape discrimination method. Using a Monte Carlo pulse simulation that added experimental noise to template pulses to account for variation in the noise environment, the WIMP acceptance of this data-quality selection was determined to be $\gtrsim$95\%.

The second level of event-selection criteria (preselection) removes event configurations inconsistent with WIMPs.  Events coincident with the muon veto are rejected ($98.7\%$ acceptance). A single-scatter requirement removes events with energy depositions in multiple detectors, a common signature for background interactions but not expected for a WIMP-nucleon scatter ($>$99\% acceptance, with losses due to noise fluctuations).  We also require events to lie within the 3$\sigma$ nuclear-recoil band and to have phonon partitions consistent with bulk nuclear recoils. A loose fiducial volume constructed from the ionization partitions further restricts events to be consistent with bulk nuclear recoils. In the radial direction, events near the detectors' sidewalls are rejected by requiring the guard electrodes on both faces to be within 2$\sigma$ from the mean of the baseline noise. For one detector (T5Z3) that has a malfunctioning guard electrode on one side, this requirement is applied on only the functioning face. A second detector (T5Z2) suffered sporadic excess noise on one guard, so only the guard on the functioning face was used for part of the dataset. In the \textit{z} direction, events on the flat faces are excluded by requiring that the inner electrodes on each side measure similar ionization energies~\cite{2013iZIPdiscrimination}. 

%--- BDT cut ---

The final level of event selection (discrimination) uses a boosted decision tree (BDT)~\cite{oai:arXiv.org:physics/0703039}.  The discriminators used by the BDT are the total phonon energy, ionization energy, phonon radial partition and phonon \textit{z} partition. Near threshold, the latter two variables provide identification of surface events superior to the ionization partitions, while the two energy quantities together optimize the discrimination at low energy where the electron-~and nuclear-recoil bands overlap. A BDT was trained for each detector using simulated background events (described below) and nuclear recoils from $^{252}$Cf calibration weighted to mimic a WIMP energy spectrum, accounting for the selection criteria acceptance. The BDT discrimination thresholds for individual detectors were chosen simultaneously to minimize the expected 90\% confidence level (C.L.) Poisson upper limit of the rate of passing events per WIMP exposure.   The BDT was trained and optimized separately for 5, 7, 10, and 15~\gev~WIMPs.  Events that pass any of the four WIMP-mass optimizations are accepted into the signal region as candidates.  When a limit is set using the optimum interval method \cite{upper, 2007arXiv0709.2701Y}, this acceptance technique provides sensitivity to a range of masses, but incurs only modest sensitivity loss compared to an analysis optimized at every WIMP mass.  In addition to the BDT, two other discrimination methods were developed and similarly optimized for WIMP masses between 5 and 15~\gev.   The BDT was chosen as the primary discrimination method before unblinding because of its better expected sensitivity on the background simulation data. 

The acceptance of the preselection criteria and the BDT was evaluated using the fraction of $^{252}$Cf nuclear recoils passing as a function of energy. Unlike WIMPs, the $^{252}$Cf neutrons can multiply scatter within a single detector, which necessitates correcting the acceptance upwards by $\sim$20\% above $\sim$5~keV$_{\rm nr}$ based on a Geant4~\cite{Agostinelli:2002hh} neutron simulation, which includes constraints on the resolution effects and the size of the fiducial volume. The uncertainty of the total acceptance is dominated by systematic uncertainty on the size of the fiducial volume and is shown in Fig.~\ref{efficiency}.

%--- backgrounds  ----

\begin{figure}[t!!]
\begin{center}
\includegraphics[width=250 pt, trim =5 0 0 5, clip=true]{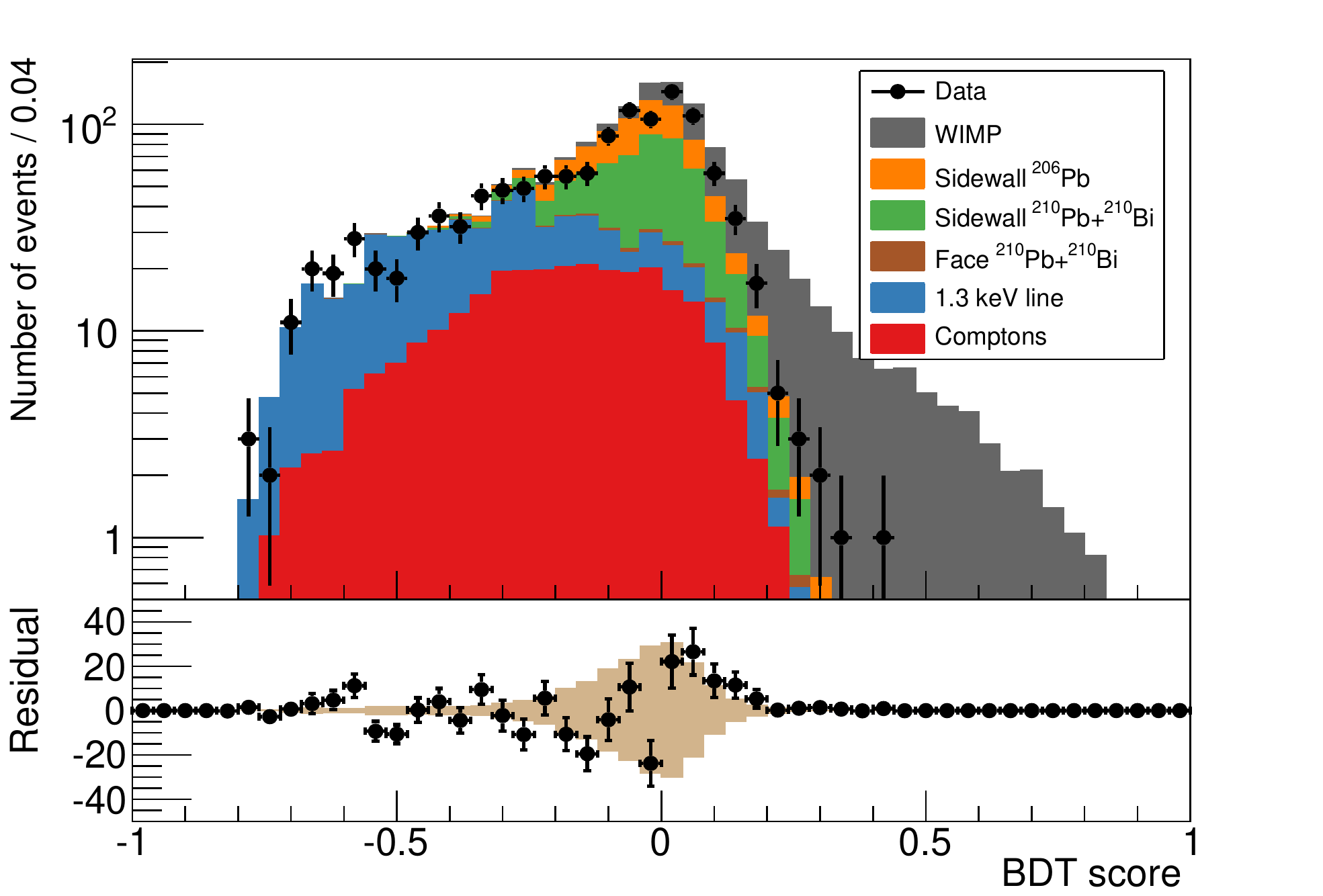}
\caption{Top: Stacked histogram showing the components of the background model passing the preselection criteria, summed over all detectors (neutron backgrounds are negligible and not included).  For comparison, a 10~\gev~WIMP with cross section $6\times10^{-42}$~cm$^2$ is shown on top of the total background.  Events passing preselection criteria are overlaid (markers with statistical errors).  A p-value statistic comparing the data to background model is 14\% for this selection. Bottom: Difference between the data and the background expectation.  Tan bars indicate the systematic uncertainty (68\% C.L.) on the background estimate.  The background spectrum was computed prior to unblinding and was not fit or rescaled to match the data.}
\label{bdt_sigback}
\end{center}
\end{figure}

A background model was developed that includes Compton recoils from the gamma-ray background; ~1.1--1.3 keV X-rays and Auger electrons from L-shell electron-capture (EC) decay of $^{65}$Zn, $^{68}$Ga, $^{68}$Ge and $^{71}$Ge; and decay products from $^{210}$Pb contamination on the detectors and their copper housings. We normalize the flat Compton background to the observed rate of electron recoils in the range 2.6--5.1 keV$_{\rm ee}$. The average rate of L-shell EC events is estimated by scaling the observed rate in the open dataset by the ratio of the K-shell event rates in the WIMP-search and open datasets. We use Geant4 to simulate the implantation and decay of $^{222}$Rn daughters starting from $^{214}$Po as described in~\cite{2013iZIPdiscrimination}. Background components from $^{210}$Pb decay products (betas, conversion electrons, X-rays), $^{210}$Bi betas, and $^{206}$Pb nuclei from $^{210}$Po decays are considered, with rates normalized to the alpha and $^{206}$Pb decay products of $^{210}$Po under the assumption of secular equilibrium.

The background model is implemented using events from high-energy sidebands and calibration data as templates for low-energy backgrounds. Ionization and phonon pulses are scaled to lower energies, injected with noise from randomly triggered events throughout the data, and reconstructed as actual data.  $^{133}$Ba calibration data and K-shell EC events are used as templates for the Compton recoils and L-shell EC events, respectively. Templates for $^{210}$Pb daughters are sampled from high-energy betas and $^{206}$Pb recoils.

%----- box opening and discussion

Figure \ref{bdt_sigback} shows the individual components of the background model as a function of the 10~\gev\ BDT discrimination parameter after applying the preselection criteria.  This background model was finalized prior to unblinding and predicted 6.1$^{+1.1}_{-0.8}$ (stat.+syst.) events passing the BDT selection. Simulations of radiogenic and cosmogenic neutrons, as described in~\cite{CDMSScience:2010}, predict an additional $0.098 \pm 0.015$~(stat.) events.   These estimates included only known systematic effects. Because the accuracy in background modeling required for a full likelihood analysis is difficult to achieve in a blind analysis of this type, the decision was made before unblinding to report an upper limit on the WIMP-nucleon cross section

\begin{figure}[t!!]
\begin{center}
\includegraphics[width=250 pt]{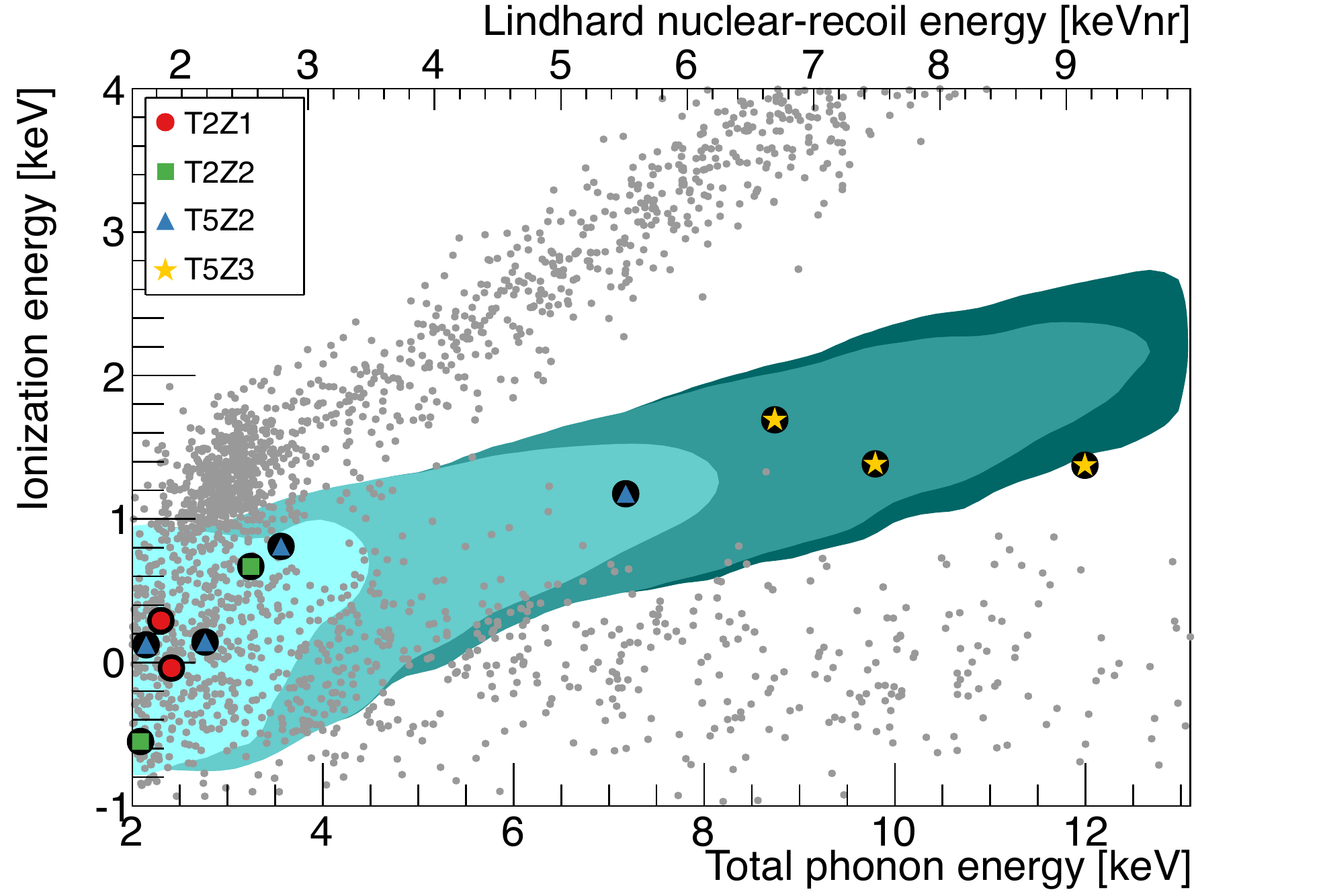}
\caption{Small gray dots are all veto-anticoincident single-scatter events within the ionization-partition fiducial volume that pass the data-quality selection criteria. Large encircled shapes are the 11 candidate events.  Overlapping shaded regions (from light to dark) are the 95\% confidence contours expected for 5, 7, 10 and 15~\gev~WIMPs, after application of all selection criteria.  The three highest-energy events occur on detector T5Z3, which has a shorted ionization guard.  The band of events above the expected signal contours corresponds to bulk electron recoils, including the 1.3~keV activation line at a total phonon energy of $\sim$3~keV.  High-radius events near the detector sidewalls form the wide band of events with near-zero ionization energy.  For illustrative purposes, an approximate nuclear-recoil energy scale is provided.}
%\caption{Gray markers are triggered single-scatter events passing all basic data quality criteria.  The large dots are candidates selected by the BDT.  The shaded regions are the 95\% C.L. regions expected for WIMPs of mass 5, 7, 10 and 15~\gev, after the BDT selection criteria are applied.  The three highest energy events occur on an impaired detector (see text discussion). Above the signal bands is the bulk electron band including the 1.3 keV activation line at a total phonon energy of $\sim$3~keV. Outer radial events form the wide horizontal band around zero ionization energy.}
\label{candidates_plot}
\end{center}
\end{figure}

Upon unblinding, eleven candidates were observed as indicated in Fig.~\ref{candidates_plot}. The events were found to be of high quality and occurring during good periods of experimental operation, except for the lowest-energy candidate, which has an abnormal pulse shape and is suspected to be noise.   As seen in Table~\ref{candidates_table}, the observed number of events is consistent with the background prediction for most detectors. However, the three high-energy events in detector T5Z3 strongly disagree with the background prediction.  The probability to observe at least this many background events on this detector is $4 \times 10^{-4}$.  These events are observed on the only detector in this dataset that has an ionization guard electrode shorted to ground.  Although the background model was developed to account for the shorted channel, we realized after unblinding that the altered electric field may have affected the selection of background model templates, potentially making the background estimate on this detector inaccurate.

The background model is compared to unblinded events passing all preselection criteria in Fig.~\ref{bdt_sigback}.  The systematic uncertainty, shown with tan fill, is dominated by the uncertainty of the expected ionization of sidewall events originating from $^{210}$Pb and $^{210}$Bi.  P-value statistics comparing the data passing the preselection criteria with the blind background model prediction range from 8--26\% for the BDTs trained to each of the four WIMP masses. This reasonable compatibility, based on the sum over all detectors, suggests that the background model correctly reproduces most features of the true background.

\begin{table}[b!]
\begin{center}
\begin{tabularx}{225pt}{ X  X  X }
\hline\hline                       
         & Candidate              &  Expected \\
Detector & energies [keV$_{\rm nr}$] &  background\\\hline
T1Z1 & ---                & 0.03$^{+0.01}_{-0.01}$ \\
T2Z1 & 1.7, 1.8           & 1.4$^{+0.2}_{-0.2}$ \\
T2Z2 & 1.9, 2.7           & 1.8$^{+0.4}_{-0.3}$ \\
T4Z2 & ---                & 0.04$^{+0.02}_{-0.02}$ \\
T4Z3 & ---                & 1.7$^{+0.4}_{-0.3}$ \\
T5Z2 & 1.9, 2.3, 3.0, 5.8 & 1.1$^{+0.3}_{-0.3}$ \\
T5Z3 & 7.0, 7.8, 9.4      & 0.13$^{+0.06}_{-0.04}$ \\  % T5Z3 still needs to be updated!
\hline\hline
\end{tabularx}
\caption{Energies of candidate events in each detector, labeled by tower (first number) and position within tower from top to bottom (second number).  Expected background is based on the model used to train the BDT and includes the estimated systematic uncertainty. Differences in expected background across detectors reflect different trigger thresholds and background event rates. Event energies are calculated using the measured mean ionization energy for nuclear recoils.}
\label{candidates_table}
\end{center}
\end{table}

\begin{figure}[t!]
\begin{center}
\includegraphics[width=250 pt]{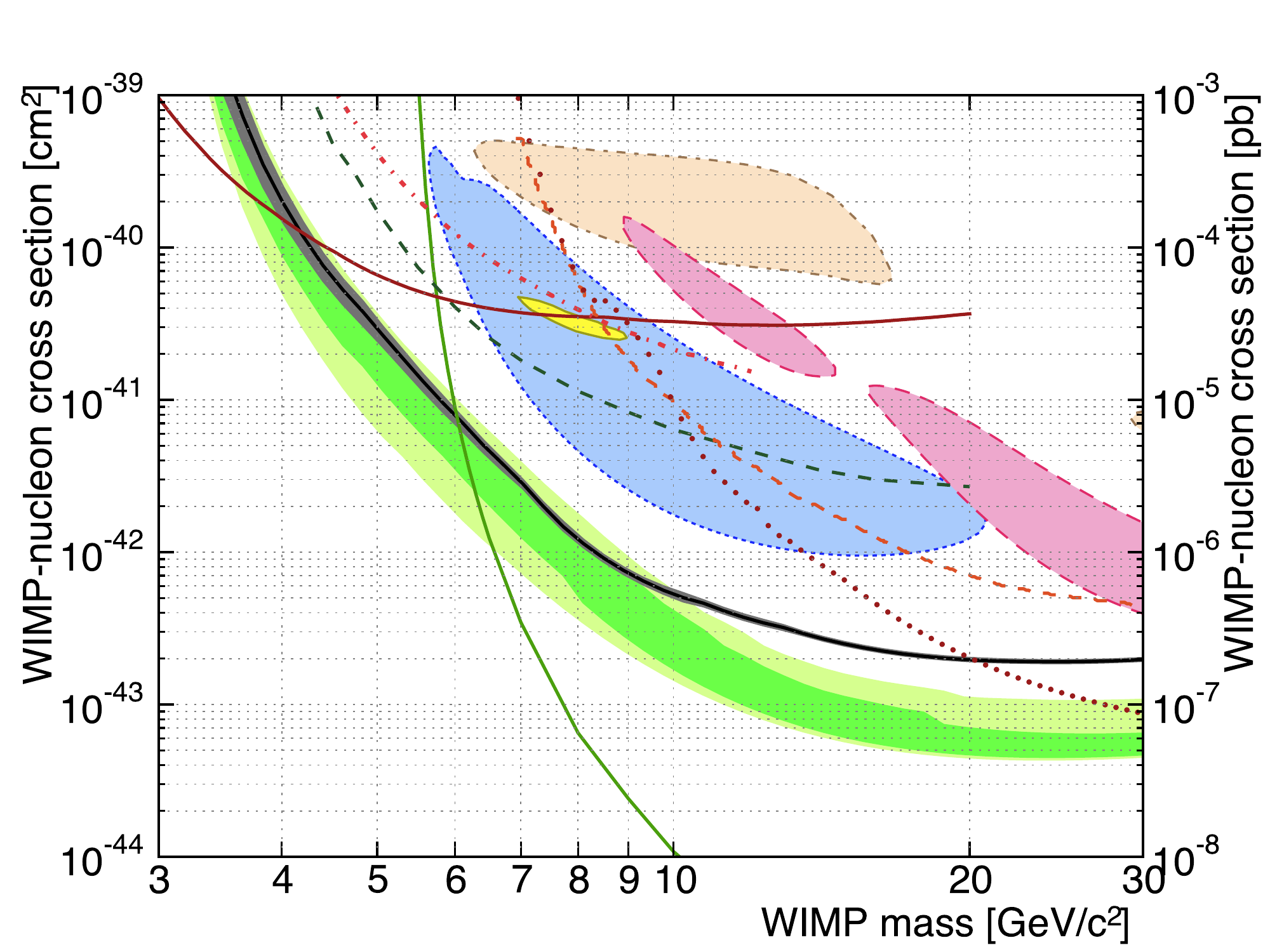}
\caption{The 90\% confidence upper limit (solid black) based on all observed events is shown with 95\% C.L. systematic uncertainty band (gray).  The pre-unblinding expected sensitivity in the absence of a signal is shown as 68\% (dark green) and 95\% (light green) C.L. bands.  The disagreement between the limit and sensitivity at high WIMP mass is due to the events in T5Z3.  Closed contours shown are CDMS II Si \cite{Agnese:2013rvf} (\emph{dotted blue}, 90\% C.L.), CoGeNT \cite{PhysRevD.88.012002} (\emph{yellow}, 90\% C.L.), CRESST-II \cite{Angloher:2012fk} (\emph{dashed pink}, 95\% C.L.), and DAMA/LIBRA \cite{Savage:2008er} (\emph{dash-dotted tan}, 90\% C.L.).  90\% C.L. exclusion limits shown are CDMS II Ge \cite{CDMSScience:2010} (\emph{dotted dark red}), CDMS II Ge low-threshold \cite{PhysRevLett.106.131302} (\emph{dashed-dotted red}), CDMSlite \cite{cdmslitePRL} (\emph{solid dark red}), LUX \cite{PhysRevLett.112.091303} (\emph{solid green}), XENON10 S2-only \cite{PhysRevLett.107.051301,PhysRevLett.110.249901} (\emph{dashed dark green}), and EDELWEISS low-threshold \cite{PhysRevD.86.051701} (\emph{dashed orange}). 
} 
\label{limitplot}
\end{center}
\end{figure}

A 90\% C.L. upper limit on the spin-independent WIMP-nucleon cross section was calculated using the optimum interval method without background subtraction. The calculation used standard halo assumptions as discussed in~\cite{inelastic}. The result is shown in Fig.~\ref{limitplot}.  Statistical and systematic uncertainties in the fiducial-volume efficiency, the nuclear-recoil energy scale, and the trigger efficiency were propagated into the limit by Monte Carlo and are represented by the narrow gray band around the limit.  The limit is consistent with the expected sensitivity for masses below 10~\gev\, as shown by the green band in Fig.~\ref{limitplot}. The discrepancy above 10~\gev\ is due to the three high-energy events in T5Z3, which are in tension with the background expectation.

This work represents the first search for WIMPs with the background rejection capability of SuperCDMS detectors. A physically motivated background model generally agrees with the data, except for the detector with a shorted ionization guard. This analysis strongly disfavors a WIMP-nucleon scattering interpretation of the excess reported by CoGeNT, which also uses a germanium target. Similar tension exists with WIMP interpretations of several other experiments, including CDMS~II (Si), assuming spin-independent interactions and a standard halo model.  New regions of WIMP-nucleon scattering for WIMP masses below 6~\gev\ are excluded.

%--- the end ---

%depracated
%\input{introduction}
%\input{dataset}
%\input{calibration}
%\input{energyscale}
%\input{qualitycuts}
%\input{preselect}
%\input{backgrounds}
%\input{PFiducialVolume}
%\input{efficiencies}
%\input{results} 
%\input{optimuminterval}

% ****** acknowledgements ****** 

%I took out the grant number, which traditionally have appeared here.  My understanding from Tarek is that we no longer have to include these.  We should also check on whether Dennis Seits, Larry Novak and Bruce Hines should be explicitly called out here.

The SuperCDMS collaboration gratefully acknowledges the contributions of numerous engineers and technicians.  In addition, we gratefully acknowledge assistance from the staff of the Soudan Underground Laboratory and the Minnesota Department of Natural Resources. The iZIP detectors were fabricated in the Stanford Nanofabrication Facility, which is a member of the National Nanofabrication Infrastructure Network. This work is supported in part by the National Science Foundation, by the United States Department of Energy, by NSERC Canada, and by MultiDark (Spanish MINECO). Fermilab is operated by the Fermi Research Alliance, LLC under Contract No. De-AC02-07CH11359. SLAC is operated under Contract No. DE-AC02-76SF00515 with the United States Department of Energy.

% ****** references ****** 

\bibliographystyle{apsrev4-1}
\bibliography{biblio_R133_LowT_PRL}

\end{document}